# Atomic adsorption on pristine graphene along the Periodic Table of Elements – From PBE to non-local functionals

Igor A. Pašti[1] *, Aleksandar Jovanović[1,2], Ana S. Dobrota[1], Slavko V. Mentus[1,3], Börje Johansson[4], Natalia V. Skorodumova[4,5]

[1]*University of Belgrade – Faculty of Physical Chemistry, Studentski trg 12-16, 11158 Belgrade, Serbia*

[2]*CEST Kompetenzzentrum für elektrochemische Oberflächentechnologie GmbH, Viktor-Kaplan-strasse 2, Section A, 2700 Wiener Neustadt, Austria*

[3]*Serbian Academy of Sciences and Arts, Knez Mihajlova 35, 11000 Belgrade, Serbia*

[4]*Department of Physics and Astronomy, Uppsala University, Box 516, 751 20 Uppsala, Sweden*

[5]*Department of Materials Science and Engineering, School of Industrial Engineering and Management, KTH - Royal Institute of Technology, Brinellvägen 23, 100 44 Stockholm, Sweden*

* **corresponding author**, e-mail: igor@ffh.bg.ac.rs

**The understanding of atomic adsorption on graphene is of high importance for many advanced technologies. Here we present a complete database of the atomic adsorption energies for the elements of the Periodic Table up to the atomic number 86 (excluding lanthanides) on pristine graphene. The energies have been calculated using the projector augmented wave (PAW) method with PBE, long-range dispersion interaction corrected PBE (PBE+D2, PBE+D3) as well as non-local vdW-DF2 approach. The inclusion of dispersion interactions leads to an exothermic adsorption for all the investigated elements. Dispersion interactions are found to be of particular importance for the adsorption of low atomic weight earth alkaline metals, coinage and s-metals ($11^{th}$ and $12^{th}$ groups), high atomic weight p-elements and noble gases. We discuss the observed adsorption trends along the groups and rows of the Periodic Table as well some computational aspects of modelling atomic adsorption on graphene.**

# 1. Introduction

Due to its extraordinary properties, graphene is constantly in the research focus of contemporary materials science. Unique electronic structure, combined with high surface area, exceptional thermal and mechanical properties made graphene attractive for many applications [1,2]. It is also known to be chemically inert, only weakly interacting with other chemical species [3,4]. However, better understanding of graphene reactivity would be useful for a variety of applications, where graphene or other carbon materials are the key elements.

Considering the huge interest in graphene, it is natural that a large number of published studies address the interactions of different atoms with graphene basal plane. However, very few systematic reports can be found. We can mention the work by Nakada and Ishii [5,6] where the atomic adsorption of the elements of the Periodic Table Elements (PTE) from H to Bi, excluding noble gases and lanthanides, has been investigated. In this study the spin non-polarized DFT-LDA approach is used, which is known to result in overbinding. Chan *et al*. [7] have addressed the adsorption of 12 metals on graphene basal plane. In addition, Valencia *et al*. [8] have analyzed the adsorption of 3d elements on graphene using periodic GGA PW91 calculations and compared the obtained adsorption trends with that on carbon nanotubes. The most recent work is the review paper by Widjaja *et al*. [9] which provides the results for atomic adsorption on graphene basal plane obtained using spin polarized DFT calculations with dispersion correction in the formulation of Grimme [10]. The authors also addressed the magnetism of the adsorbate+graphene (A@G) systems. Naturally, due to different computational schemes applied, the estimated adsorption energies differ significantly in some cases but here we focus on the overall trends. Even taken together these reports do not give a complete overview of the trends in the atomic adsorption of the elements of the PTE. In fact, opposing trends are observed in many cases. For example, the reports of Nakada and Ishii [5,6] predict that the strongest adsorption of the elements in the d-block of the PTE is for the elements with the half filled shell. In contrast, Valencia *et al*. [8] predict that such elements in the 3d series (Cr and Mn) only weakly physisorb on graphene. While the binding energy maxima are also expected along the 4d and 5d series based on the LDA studies [5,6], Widjaja *et al*. [9] predict different trends along the 3d series, on one hand, and along 4d and 5d series, on the other. The same holds for the p-elements. First, strong bonding is reported for the elements in the middle of p-series [6]. However, afterwards weak bonding of $np^3$ elements is predicted using dispersion-corrected GGA [9]. Moreover, previous LDA studies [5,6] predict rather strong adsorption for some elements, like in the case of O, for which the LDA binding energy is twice as large as the GGA one [9]. Moreover, the work of Widjaja *et al*. [9] claims unstable adsorption for most of the elements in the 6$^{th}$ row of the PTE (from Cs to Bi) that contradicts numerous reports (for example, see Refs. [4,11] and references therein). Additionally, the chemical inertness of graphene frequently invokes the question of the necessity of the inclusion of dispersion interaction in modelling of adsorption on graphene. Based on the results overviewed above, it is not possible to conclude for which elements it is necessary to include the dispersion interactions. Also, it cannot be estimated how the use of different schemes for the treatment of the dispersion interactions affects the results.

Here we present a detailed, comprehensive view on the atomic adsorption on the graphene basal plane for the elements of the PTE providing benchmark values for the adsorption energies of atoms on pristine graphene, which can be further transferred to other types of carbon materials [8]. To be able to do this we performed carefully converged density functional theory (DFT) calculations using different computational schemes – from common PBE to dispersion-corrected PBE and non-local DFT calculations. The calculations were carried out for all the elements in rows 1-6 of the PTE except lanthanides. The presented results enable us to analyze the trends in atomic adsorption on graphene and to address the importance of the explicit treatment of dispersion interactions. Moreover, we address some computational issues, such as the number of valence electrons, and compare the results of the two common DFT codes – VASP and Quantum ESPRESSO.

## 2. Computational details

We calculated the adsorption of all the elements of the PTE located in rows 1 to 6 (except lanthanides), on pristine graphene modelled as 4×4 cell (32 atoms). We chose to use a larger simulation cell compared to those used in the previous works [6,9] in order to minimize the adatom-image interaction. The repeated graphene sheets were separated from each other by 20 Å of vacuum.

The first-principle DFT calculations were performed using the Vienna *ab initio* simulation code (VASP) [12-15]. In the first step we used the generalized gradient approximation (GGA) in the parametrization by Perdew, Burk and Ernzerhof [16] and the projector augmented wave (PAW) method [17,18]. We used the PAW potentials provided with the latest VASP distribution and, when available, those which include the semi-core states into the valence band. A cut-off energy of 600 eV and Gaussian smearing with a width of $\sigma = 0.025$ eV for the occupation of the electronic levels were used. A Monkhorst-Pack $\Gamma$-centered 10×10×1 k-point mesh was used. Three possible adsorption sites were investigated (top, bridge and hollow) and during the optimization the relaxation of adatoms was allowed only in the direction vertical to the graphene basal plane. At the same time, the relaxation of all carbon atoms was unrestricted. The relaxation procedure was stopped when the Hellmann-Feynman forces on all atoms were smaller than $10^{-2}$ eV Å$^{-1}$. This corresponded to the total energy converged below 0.01 meV. Spin-polarization was taken into account in all calculations.

To account for dispersion interactions we used different approaches. First, we used DFT theory plus long-range dispersion correction in the DFT+D2 and DFT+D3 formulations of Grimme [10,19]. Both approaches correct the total energy by a pairwise term, which accounts for dispersion interactions and which is added to the total energy of the system calculated using a selected DFT functional (in this case PBE):

$$E_{PBE+D} = E_{DFT\text{-}PBE} + E_{disp} \qquad (1)$$

$E_{disp}$ term depends on the pair of interacting atoms. It is obtained by the summation of atom-specific parameters and relative distances over the entire simulation cell. Within the DFT+D2 formulation these parameters are insensitive to chemical surrounding. In the DFT+D3 scheme the parameters are geometry-dependent and are adjusted for local geometry. For DFT+D2 we used the default set of parameters (as implemented in VASP) for the elements in rows 1-5. For the elements of the 6$^{th}$ row we used DFT+D2 parameters as described in Ref. [20].

Additionally, we applied the vdW-DF2 non-local functional developed by Langreth's and Lundqvist's groups [21]. The method relies on the ideas of Dion *et al*. [22] and Roman-Perez and Soler [23] and it is implemented in VASP in a way allowing for the inclusion of the non-local contribution into the correlation energy during the self-consistency cycle [24]. The accuracy of non-local functionals is known to generally increase when the number of valence electrons taken into consideration increases. To explicitly check this we compared the performance of different PAW datasets for a number of selected atoms.

The adsorption energies obtained within different computational schemes were calculated as:

$$E_{ads}^{PBE} = E_0^{PBE}[A@G] - E_0^{PBE}[G] - E_0^{PBE}[A] \quad (2)$$

$$E_{ads}^{PBE+D} = E_0^{PBE+D}[A@G] - E_0^{PBE+D}[G] - E_0^{PBE}[A] \quad (3)$$

$$E_{ads}^{vdW-DF2} = E_0^{vdW-DF2}[A@G] - E_0^{vdW-DF2}[G] - E_0^{vdW-DF2}[A] \quad (4)$$

where $E_0$ are the ground state energies of the adatom on graphene [A@G], graphene [G] and adatom [A] alone, calculated with the specified method. $E_{ads}$ is negative when adsorption is exothermic.

In addition, for a selected set of atoms we performed calculations by using both VASP and Quantum ESPRESSO code [25] and compared the obtained results. For the calculations with Quantum ESPRESSO ultrasoft pseudopotentials, PBE and PBE+D2 functionals were used.

## 3. Results and discussion

### 3.1. Graphene sheet

The optimized C–C bond lengths (Table 1) were calculated by fitting the total energy *vs.* C–C bond length curves (Supplementary Information, Fig.S1). Calculated values are in good agreement with the experimental C–C bond length in graphene sheet (1.421 Å [26]). The calculated cohesive energies of graphene (Table 1) can be compared to the results of quantum Monte Carlo study of Shin *et al.* [27]. The authors reported the value of the cohesive energy of graphene to be −7.972 eV atom$^{-1}$ as calculated using the PBE-PAW approach. Other previously reported results obtained using the same approach are −7.73 eV atom$^{-1}$ [28] and −7.9 eV atom$^{-1}$ [27]. The result of Shin *et al*. [27] for the cohesive energy of graphene, obtained using the zero-point energy-corrected diffusion Monte Carlo method, amounts to −7.298 eV atom$^{-1}$. This value is very close to our vdW-DF2 result, while our PBE and dispersion corrected PBE calculations give the results in close agreement with the previously reported PBE-PAW values (Table 1). In order to estimate the experimental cohesive energy of graphene one can use its interlayer cohesive energy, which is estimated [29] to be −0.0615 eV atom$^{-1}$ and the experimentally determined cohesive energy of graphite, −7.374 eV atom$^{-1}$ [27]. Hence, the experimental estimate of graphene cohesive energy is around −7.312 eV atom$^{-1}$, which is very close to our vdW-DF2 result (Table 1).

**Table 1.** Optimized C–C bond lengths and calculated cohesive energies of graphene for the applied computational schemes

| Computational scheme | C–C bond length / Å | Cohesive energy* / eV atom$^{-1}$ |
|---|---|---|
| PBE | 1.425 | −7.828 |
| PBE+D2 | 1.425 | −7.883 |
| PBE+D3 | 1.425 | −7.863 |
| vdW-DF2 | 1.430 | −7.268 |

*calculated as $E_0^X[\text{G}]/n - E_0^X[\text{C}]$, where X stands for the computational scheme and $n = 32$.

The analysis of the calculated densities of states (DOS) of pristine graphene indicates that all the four computational schemes adequately reproduce the electronic structure of graphene (for details see Supplementary Information, Fig. S2).

Good agreement between our results and the previously reported theoretical and experimental data regarding graphene structural and electronic properties confirms that we have an adequate structural model, which we further use to study atomic adsorption on pristine graphene.

3.2. Atomic adsorption on pristine graphene

On pristine graphene one can identify three high symmetry sites available for atomic adsorption: top (single-coordinated), bridge (two-coordinated) and hollow (six-coordinated) sites. We define the preferential adsorption site as the one with the most exothermic adsorption. The identified preferential adsorption sites are presented in Table S1 (Supplementary Information), while the adsorbate-carbon distances for the preferred adsorption sites are provided in Table S2. We find that for a number of elements the energy difference between at least two adsorption sites is very small (below thermal energy at 298 K). Therefore, the mobility of these adatoms on pristine graphene is expected to be very high. These elements include low atomic weight earth alkaline metals, coinage and s-metals ($11^{th}$ and $12^{th}$ group), high atomic weight p-elements and noble gases. We note that the identified preferential adsorption sites for different computational schemes agree in many cases. However, some differences can be seen and we can conclude that vdW-DF2 functional often favors less coordinated adsorption sites, compared to PBE and dispersion corrected-PBE.

The calculated adsorption energies obtained using the four different computational schemes are provided in Figs. 1–4. The summarized adsorption energies for all four schemes are also tabulated in Table S3 (Supplementary Information).

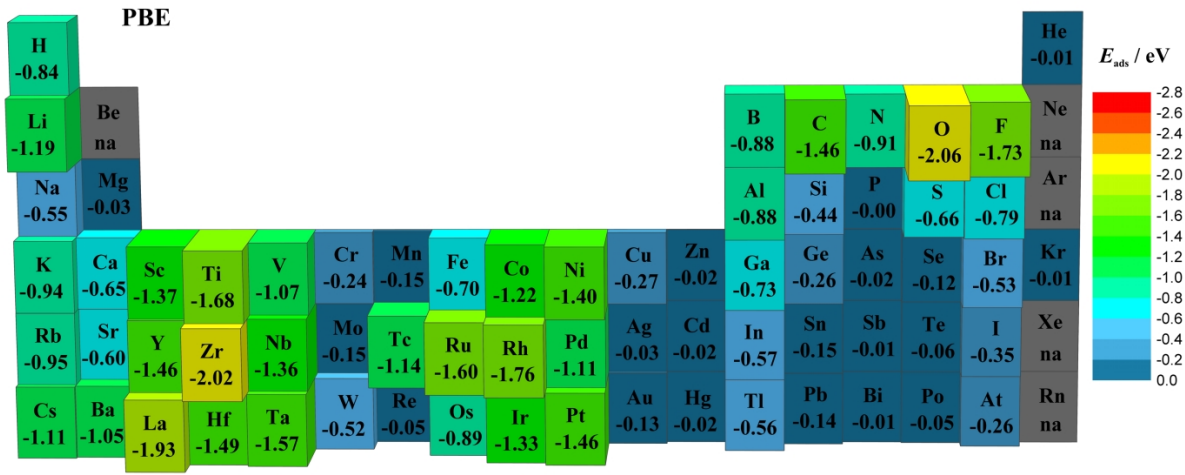

**Figure 1.** 3D plot of adsorption energies (given in eV) of investigated elements on pristine graphene obtained within the PBE scheme (na = no adsorption).

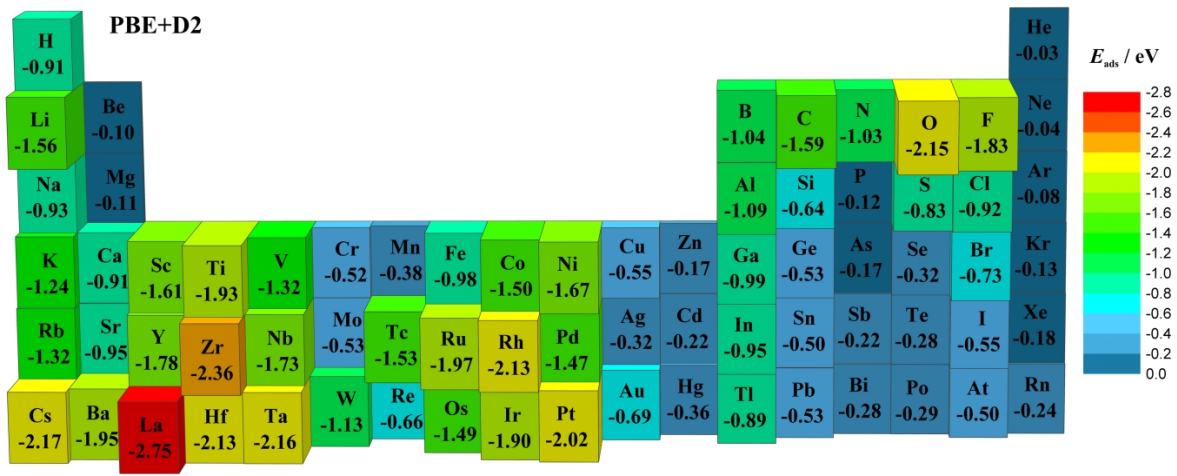

**Figure 2.** 3D plot of adsorption energies (given in eV) of investigated elements on pristine graphene obtained within the PBE+D2 scheme.

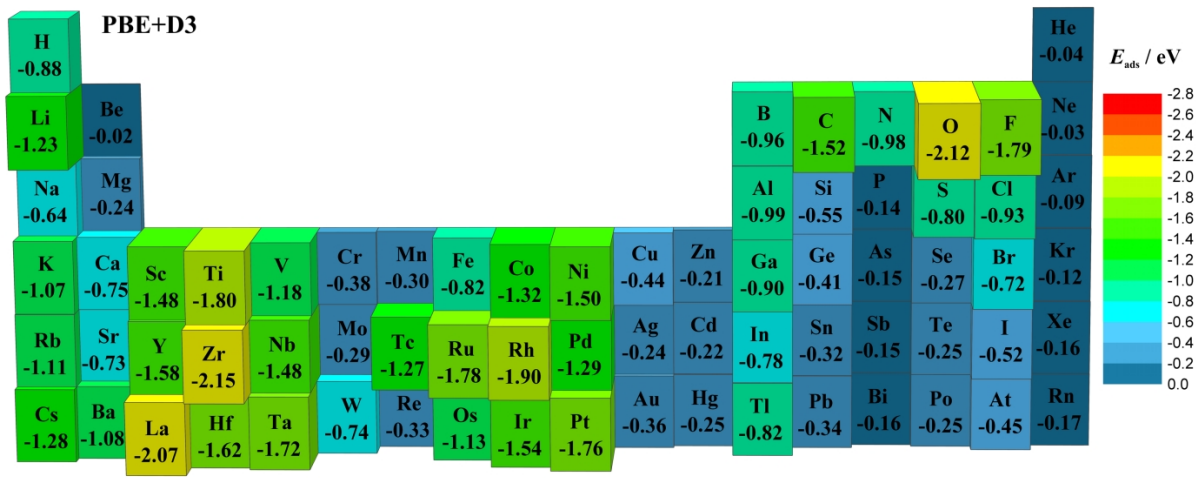

**Figure 3.** 3D plot of adsorption energies (given in eV) of investigated elements on pristine graphene obtained within the PBE+D3 scheme.

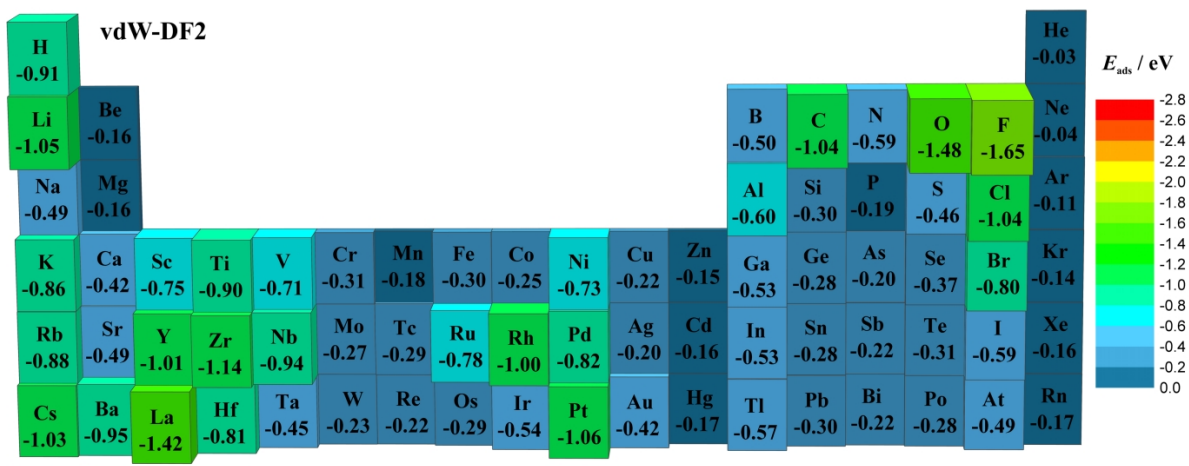

**Figure 4.** 3D plot of adsorption energies (given in eV) of investigated elements on pristine graphene obtained within the vdW-DF2 scheme.

All the four used computational schemes capture similar trend in atomic adsorption on pristine graphene (Fig. 5). The typical order of the calculated adsorption strength is vdW-DF2 < PBE < PBE+D3 < PBE+D2. This is, however, not the case for the already mentioned low atomic weight earth alkaline metals, coinage and s-metals (11th and 12th group), high atomic weight p-elements and noble gases for which PBE predicts very weak or no adsorption at all. In these cases dispersion interactions contribute significantly to calculated $E_{ads}$ and both empirical computational schemes and non-local vdW-DF2 calculations predict exothermic adsorption.

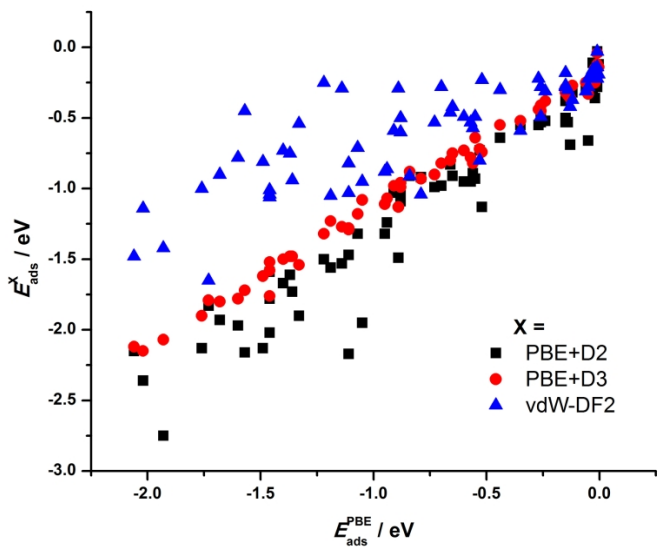

**Figure 5**. Correlation between $E_{ads}^{PBE+D2}$ (■), $E_{ads}^{PBE+D3}$ (●), $E_{ads}^{vdW-DF2}$ (▲) and $E_{ads}^{PBE}$

We find that the nature of graphene interaction with different elements of the PTE varies to a great extent. To illustrate this we use Na, O and Ne as representatives of different types of bonding with graphene. In particular, in the case of Na dominantly ionic interaction with charge transfer to Na is observed. The positively charged Na atom is centered above the carbon hexagon, the atoms of which receive the charge from the adatom. Na adsorption leads to an upshift of the Fermi level due to a charge transfer from Na to graphene without any significant disruption of the π electronic system or graphene corrugation (Fig. 6). In the case of O a covalent bond with graphene basal plane is formed. There is a significant re-hybridization of the sp$^2$ orbitals of carbon atoms surrounding the adsorption site, leading to deformation of basal plane (see Fig. 6). In the case of Ne, dispersion dominates the interaction and there is no appreciable charge transfer (Fig. 6). The adatom is rather far from the graphene basal plane, which is not deformed as, for example, in the case of O adsorption.

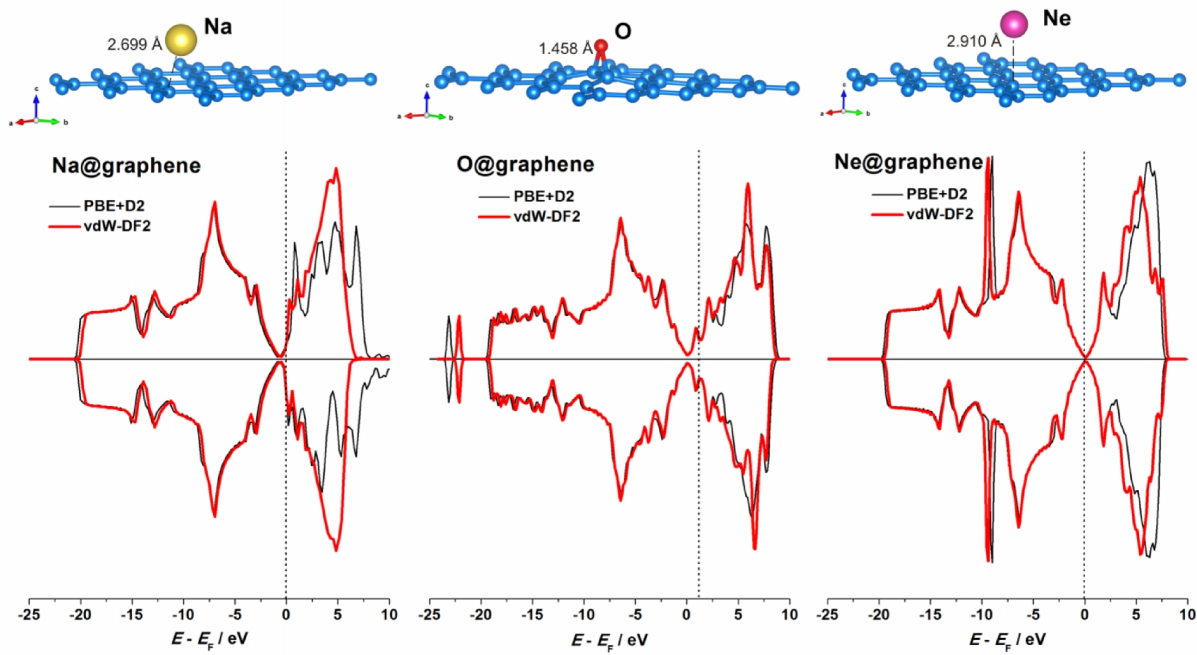

**Figure 6.** DOS plots for Na (lef), O (middle) and Ne adsorption on pristine graphene obtained using PBE+D2 scheme (thin black line) and vdW-DF2 functional (thick red line). In the upper line corresponding optimized geometries using PBE+D2 scheme are given. Indicated numbers give the adsorbate-carbon distances for each considered case.

We conclude that the use of dispersion-corrected or non-local functionals is essential when interactions are weak and no formation of true chemical bond occurs. In all the considered cases weak interaction corresponds to relatively large distances between the adatom and graphene basal plane (see structures in Fig. 6) and we conclude that in these cases the attractive part of the non-local functional dominates the interactions. When the adatom is close to the graphene basal plane, the exchange-correlation part is repulsive that reduces the energy. The vdW-DF2 scheme can be expected to give the best estimate for adsorption energy whereas PBE and dispersion-corrected PBE schemes can to result in overbinding.

One can notice some trends in adsorption along the PTE groups. In the case of p-elements, for example, we observe that binding gets weaker as the atomic number increases. In contrast, the adsorption of noble gases gets stronger as the atomic number increases. Interestingly, the studies of noble gas adsorption on graphene are rather rare. Noble gases were also omitted in previous systematic studies of atomic adsorption on graphene [6,9]. The interaction of noble gases with graphene is dominated by dispersion interaction; therefore, we assume that bonding is associated with a polarization of the noble gas electron cloud. In Fig. 7 we plot the calculated adsorption energies versus the noble gas atom polarizabilities [30-34] and observe a practically linear correlation for $E_{ads}^{PBE+D2}$. $E_{ads}^{PBE+D3}$ and $E_{ads}^{vdW\text{-}DF2}$ show certain deviations from the linear behavior for Rn but good overall trends are still observed. Such a correlation can be indeed expected based on the underlying theory of empirical long-range corrections for dispersion interactions as the dispersion coefficients depend on atomic polarizabilities [10,19]. It is interesting to observe that empirical correction produces the same trend as the explicit treatment of non-local contributions using the vdW-DF2 approach.

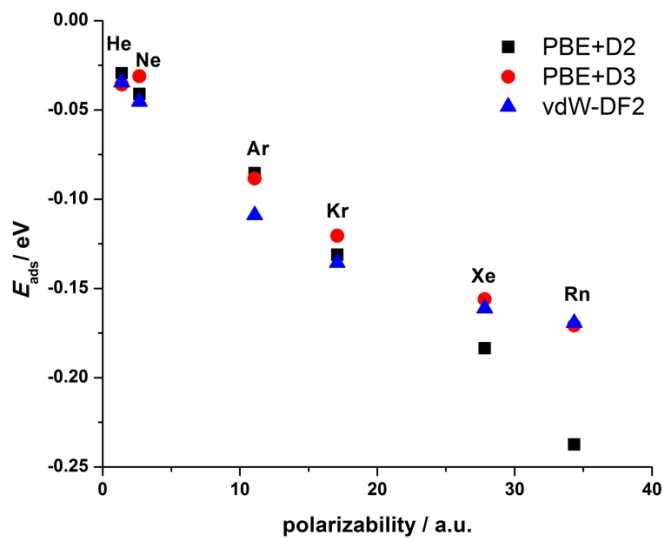

**Figure 7.** Correlation between $E_{ads}^{PBE+D2}$ (■), $E_{ads}^{PBE+D3}$ (●), $E_{ads}^{vdW-DF2}$ (▲) (given in eV) and polarizabilities of noble gas atoms. For He-Xe series experimental data were used, while for Rn theoretically obtained value is included.

One can also observe certain regularities in atomic adsorption along the rows of the PTE (see Figs. 1–4, and also Table S3, Supplementary Information). These regularities are related to the electronic configurations of the adsorbates. First, comparing alkaline and earth alkaline metals, one can see that the $ns^2$ configuration always gives weaker bonding compared to that for the $ns^1$ configuration. When the $d$ shell starts to be filled the binding strength increases, passes through a maximum and goes down for the elements with the half-filled $d$ shell (Cr, Mo, W). After that the binding strength passes through another maximum and goes down again when the $d$ shell is filled and the atoms reach the $ns^2(n-1)d^{10}$ configuration (Zn, Cd, Hg). The situation is quite similar for the p elements. Namely, the weakest bonding along the row is observed for the elements with the highest multiplicity of isolated atoms (corresponding to $np^3$ configuration). Another minimum is reached for noble gases where the electron shell is completely filled. We note that the outlined trends are visible to some extent in the work of Widjaja *et al*. [9], predominantly for 3d elements, but there are also some differences. The most pronounced difference was found for Mo, for which the strongest bonding among all the elements is predicted in Ref. [9], while we predict rather weak physisorption with a significant contribution of dispersion interactions (Figs. 1 to 4). Also, different trends are observed for the entire 6$^{th}$ row of PTE, where we predict bonding for all the elements, while Widjaja *et al*. [9] mainly predict unstable adsorption. The literature regarding particular cases of atomic adsorption on graphene is vast and we choose not to discuss it here in detail, but concentrate on the adsorption trends instead. It is important to note that the minima of the adsorption energy found for the elements with half filled and completely filled d shells were previously observed by Valencia *et al*. [8], opposing the results of Nakada and Ishii [5,6]. The authors discussed such a behavior in detail and concluded that it was due to the following factors: (i) the energy required to promote $n$s electron into the ($n$−1)d shell, (ii) metal charge donation to π* system of graphene and (iii) 4s occupation identified as the source of metal-surface Pauli repulsion [8]. The minimum of the interaction strength for Cr, Mn and Cu, being classified as physisorption, was ascribed to a high occupation of the 4s states and high metal-surface π repulsion [8]. Our results for the entire d-block of the PTE strongly suggest that such a conclusion can be generalized to all d-metals. Moreover, by adopting the same ideas, the weak physisorption of Zn, Cd and Hg can be understood on the basis of strong Pauli repulsion between the filled $n$s$^2$ orbitals and the π electronic system of graphene. Such an approach can further be extended to the elements with $n$s$^2 n$p$^3$ configurations, like nitrogen, which are known to be particularly stable. We expect that the energy gained upon adsorption on graphene cannot compensate the energy required for the rearrangement of the energy levels needed to have a bond-ready atom. Hence, these atoms remain weakly physisorbed on pristine graphene. For these elements the explicit inclusion of dispersion interactions is absolutely necessary when modelling the adsorption on pristine graphene as these interactions can exceed the PBE adsorption energies by a factor up to $10^2$ (Table S3, Supplementary Information).

The ground state magnetic moments of A@G systems are summarized in Table 2. As a rule, PBE and PBE+D2(3) schemes result in similar magnetization, while vdW-DF2 results differ in some cases.

**Table 2.** Ground state magnetizations of A@G systems (given in Bohr magnetons, $\mu_B$).

| H | | | | | element | | | | | | | | | | | | He |
|---|---|---|---|---|---|---|---|---|---|---|---|---|---|---|---|---|---|
| 0.01 | | | | | PBE | | | | | | | | | | | | 0.00 |
| 0.01 | | | | | PBE+D2 | | | | | | | | | | | | 0.00 |
| 0.01 | | | | | PBE+D3 | | | | | | | | | | | | 0.00 |
| 0.00 | | | | | vdW-DF2 | | | | | | | | | | | | 0.00 |
| **Li** | **Be** | | | | | | | | | | | **B** | **C** | **N** | **O** | **F** | **Ne** |
| 0.00 | na | | | | | | | | | | | 0.00 | 0.44 | 0.78 | 0.00 | 0.00 | na |
| 0.00 | 1.02 | | | | | | | | | | | 0.00 | 0.44 | 0.78 | 0.00 | 0.00 | 0.00 |
| 0.00 | 1.02 | | | | | | | | | | | 0.00 | 0.44 | 0.78 | 0.00 | 0.00 | 0.00 |
| 0.00 | 0.00 | | | | | | | | | | | 0.00 | 0.38 | 1.80 | 0.00 | 0.00 | 0.00 |
| **Na** | **Mg** | | | | | | | | | | | **Al** | **Si** | **P** | **S** | **Cl** | **Ar** |
| 0.25 | 0.00 | | | | | | | | | | | 0.00 | 1.42 | 3.00 | 0.00 | 0.53 | na |
| 0.23 | 0.00 | | | | | | | | | | | 0.00 | 1.43 | 3.00 | 0.00 | 0.53 | 0.00 |
| 0.25 | 0.00 | | | | | | | | | | | 0.00 | 1.43 | 3.00 | 0.00 | 0.53 | 0.00 |
| 0.36 | 0.00 | | | | | | | | | | | 0.00 | 1.85 | 2.98 | 1.77 | 0.50 | 0.00 |
| **K** | **Ca** | **Sc** | **Ti** | **V** | **Cr** | **Mn** | **Fe** | **Co** | **Ni** | **Cu** | **Zn** | **Ga** | **Ge** | **As** | **Se** | **Br** | **Kr** |
| 0.08 | 1.02 | 2.22 | 3.31 | 4.38 | 5.66 | 5.44 | 2.04 | 1.05 | 0.00 | 0.81 | 0.00 | 0.00 | 1.67 | 3.00 | 1.85 | 0.59 | 0.00 |
| 0.09 | 1.02 | 2.21 | 3.30 | 4.37 | 5.66 | 5.44 | 2.04 | 1.05 | 0.00 | 0.82 | 0.00 | 0.00 | 1.66 | 3.00 | 0.00 | 0.59 | 0.00 |
| 0.09 | 1.02 | 2.22 | 3.31 | 4.38 | 5.67 | 5.46 | 2.08 | 1.05 | 0.00 | 0.84 | 0.00 | 0.00 | 1.70 | 3.00 | 1.85 | 0.59 | 0.00 |
| 0.00 | 0.54 | 1.82 | 3.41 | 4.51 | 5.80 | 5.00 | 3.98 | 1.48 | 0.00 | 0.77 | 0.00 | 0.00 | 1.81 | 2.97 | 1.80 | 0.55 | 0.00 |
| **Rb** | **Sr** | **Y** | **Zr** | **Nb** | **Mo** | **Tc** | **Ru** | **Rh** | **Pd** | **Ag** | **Cd** | **In** | **Sn** | **Sb** | **Te** | **I** | **Xe** |
| 0.10 | 1.02 | 2.04 | 3.20 | 2.01 | 0.00 | 0.23 | 1.78 | 0.71 | 0.00 | 0.99 | 0.00 | 0.00 | 1.74 | 3.00 | 1.89 | 0.68 | na |
| 0.09 | 1.01 | 2.02 | 3.17 | 1.90 | 0.00 | 0.08 | 1.76 | 0.71 | 0.00 | 0.96 | 0.00 | 0.00 | 1.70 | 3.00 | 1.91 | 0.67 | 0.00 |
| 0.10 | 1.02 | 2.04 | 3.20 | 2.03 | 5.59 | 0.20 | 1.79 | 0.72 | 0.00 | 0.98 | 0.00 | 0.00 | 1.74 | 3.00 | 1.90 | 0.66 | 0.00 |
| 0.00 | 0.00 | 0.00 | 3.26 | 3.43 | 5.85 | 5.13 | 1.89 | 0.84 | 0.00 | 0.00 | 0.00 | 0.00 | 1.78 | 2.98 | 1.84 | 0.62 | 0.00 |
| **Cs** | **Ba** | **La** | **Hf** | **Ta** | **W** | **Re** | **Os** | **Ir** | **Pt** | **Au** | **Hg** | **Tl** | **Pb** | **Bi** | **Po** | **At** | **Rn** |
| 0.04 | 0.84 | 1.87 | 3.09 | 3.49 | 2.16 | 0.81 | 1.93 | 0.92 | 0.00 | 0.87 | 0.00 | 0.00 | 1.75 | 3.00 | 1.92 | 0.73 | na |
| 0.00 | 0.80 | 1.78 | 3.06 | 3.44 | 2.22 | 0.79 | 1.93 | 0.92 | 0.00 | 0.88 | 0.00 | 0.00 | 1.73 | 3.00 | 1.95 | 0.72 | 0.00 |
| 0.02 | 0.87 | 1.88 | 3.09 | 3.50 | 2.25 | 5.02 | 1.97 | 0.92 | 0.00 | 0.87 | 0.00 | 0.00 | 1.75 | 3.00 | 1.94 | 0.72 | 0.00 |
| 0.00 | 0.00 | 0.71 | 0.95 | 2.99 | 4.51 | 5.00 | 3.87 | 1.01 | 0.00 | 0.00 | 0.00 | 0.00 | 1.75 | 2.99 | 1.87 | 0.66 | 0.00 |

It is important to notice that large differences in the predicted ground state magnetizations of A@G systems (such as the ones for Mo and Tc, or W, Re and Os) are due to different preferential adsorption sites, predicted using different computational schemes. Also, these are the elements which rather weakly interact with pristine graphene and for which the inclusion of dispersion interactions significantly affects the description of the systems significantly (especially for Mo and Re). Hence, we suggest that the explicit treatment of dispersion interactions is also important for the prediction of magnetic properties and not only accurate description of adsorption energetics. Our results are in good agreements with the previous ones [8,9]. The observed slight differences are most likely due to the different size of the simulation cell used in Ref. [9] as compared to our parameters. We also observe that the elements with the $ns^2np^3$ configuration retain the magnetic moments of the isolated atoms, except for N where a stronger interaction with pristine graphene is observed. Such a behavior can be explained by the weak interactions of these atoms with pristine graphene and the absence of chemical bonds between them.

We also investigated some computational aspects of modelling atomic adsorption on graphene. For the set of 7 elements taken from different parts of the PTE (Table S4, Supplementary Information) we analyzed the effect of the number of electrons included into the valence on the resulting energy. In all the cases the same adsorption sites were identified as preferential ones and calculated $E_{ads}$ agreed very well. A relatively large difference (0.13 eV) was observed only for Fe with the semicore 3p states in the case of the vdW-DF2 scheme. In addition, we compared the results obtained using VASP with those obtained using Quantum ESPRESSO (QE), where ultrasoft pseudopotentials were used (PBE and PBE+D2 calculations and some of the results previously reported by us [4,35], Table 3). For the selected set of elements the same overall trends in atomic adsorption are observed, while some differences can be outlined. In the cases when there are small variations of $E_{ads}$ between different adsorption sites (like Cl, Cu and Ag; Table S3, Supplementary Information) the adsorption sites identified using PBE differ for those obtained by VASP and QE. In the cases of O, Cl, Rh and Ir we observed significant differences in calculated $E_{ds}$ (up to 0.3 eV) while the ground state magnetizations obtained using QE for the selected elements agree with the ones reported in Table 2.

**Table 3**. PBE results for atomic adsorption on pristine graphene obtained using Quantum ESPRESSO (using ultrasoft pseudopotentials) and VASP (using PAW potentials). First row gives PBE results, and the second row for each element gives PBE+D2 result (t – top site, b – bridge site, h – hollow site).

| | Quantum ESPRESSO | | | VASP | | |
|---|---|---|---|---|---|---|
| element | valence e | ads. site | $E_{ads}$ / eV | valence e | ads. site | $E_{ads}$ / eV |
| H[a] | 1 | t | −0.89 | 1 | t | −0.84 |
| | | / | / | | t | −0.91 |
| Na[b] | 1 | h | −0.45 | 7 | h | −0.55 |
| | | h | −0.81 | | h | −0.93 |
| O[c] | 6 | b | −1.83 | 6 | b | −2.06 |
| | | b | −1.93 | | b | −2.16 |
| Cl[a] | 7 | t | −0.98 | 7 | b | −0.79 |
| | | / | / | | b | −0.92 |
| Fe | 16 | h | −0.63 | 8 | h | −0.70 |
| | | h | −0.89 | | | |
| | 8 | h | −0.68 | | h | −0.98 |
| | | h | −0.96 | | | |
| Ni | 10 | h | −1.44 | 10 | h | −1.40 |
| | | h | −1.71 | | h | −1.67 |
| Cu | 11 | t | −0.26 | 11 | b | −0.27 |
| | | t | −0.53 | | t | −0.55 |
| Ru | 8 | h | −1.59 | 14 | h | −1.60 |
| | | h | −1.96 | | h | −1.97 |
| Rh | 9 | h | −1.43 | 15 | h | −1.76 |
| | | h | −1.80 | | h | −2.13 |
| Pd | 10 | b | −1.08 | 10 | b | −1.11 |
| | | b | −1.45 | | b | −1.47 |
| Ag | 11 | h | −0.03 | 11 | b | −0.03 |
| | | h | −0.33 | | h | −0.32 |
| Ir | 9 | b | −1.14 | 9 | b | −1.33 |
| | | b | −1.70 | | b | −1.90 |
| Pt | 10 | b | −1.45 | 10 | b | −1.46 |
| | | b | −2.01 | | b | −2.02 |
| Au | 11 | t | −0.14 | 11 | t | −0.13 |
| | | t | −0.70 | | t | −0.69 |

[a]reference [4]; [b]reference [35] 54 atoms cell.

## 4. Conclusions

We provide a complete computational data base for the atomic adsorption on pristine graphene. Adsorption energies were calculated using PBE, long-range dispersion interaction-corrected PBE (PBE+D2 and PBE+D3) and non-local vdW-DF2 functional for all elements in rows 1 to 6 of the PTE (excluding lanthanides). Low atomic weight earth alkaline metals, coinage as s-metal (11$^{th}$ and 12$^{th}$ group of the PTE), high atomic weight p-elements and noble gases interact weakly with pristine graphene while their mobility over graphene basal plane is expected to be very high. We conclude that for these elements the inclusion of dispersion interactions is necessary to analyze the adsorption on graphene. The adsorption energies of noble gases were found to correlate with their atomic polarizabilities. The strength of the interaction along the rows of PTE depends on the electronic configuration: it has minima for the atoms with filled shells ($n$s$^2$, $n$s$^2(n-1)$d$^{10}$ and $n$p$^6$ configurations) and atoms with half-filled shells (($n-1$)d$^5$ and $n$p$^3$ configurations) and this rule of thumb extends to the entire PTE.

**Acknowledgements**

This work was supported by the Swedish Research Links initiative of the Swedish Research Council (348-2012-6196). N.V.S. acknowledges the support provided by Swedish Research Council through the project No. 2014-5993. The computations were performed on resources provided by the Swedish National Infrastructure for Computing (SNIC) at National Supercomputer Centre (NSC) at Linköping University. I.A.P., A.S.D. and S.V.M. acknowledge the support provided by the Serbian Ministry of Education, Science and Technological Development through the project III45014. Financial support provided through the NATO Project EAP.SFPP 984925 - "DURAPEM - Novel Materials for Durable Proton Exchange

Membrane Fuel Cells" is also acknowledged. We also acknowledge the support from Carl Tryggers Foundation for Scientific Research.

**SUPPLEMENTARY INFORMATION**

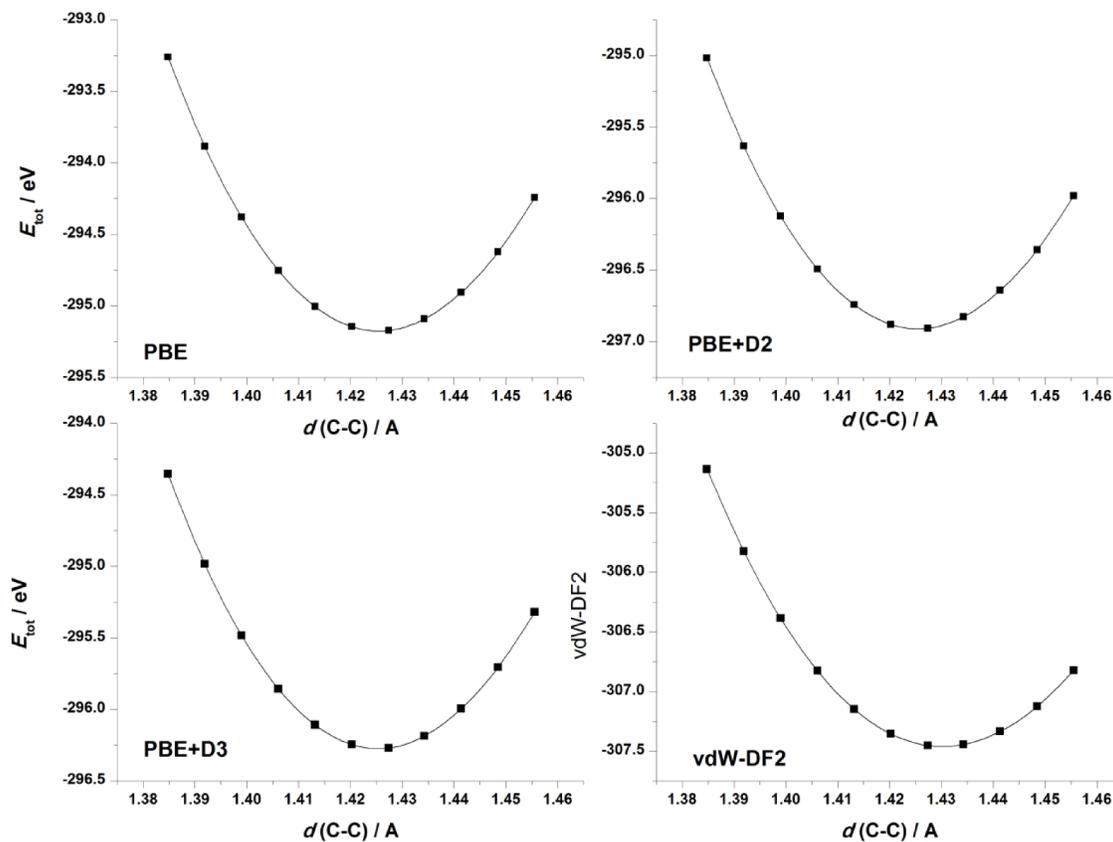

**Figure S1.** Total energy *vs*. C–C bond length curves for four different computational schemes applied in this work.

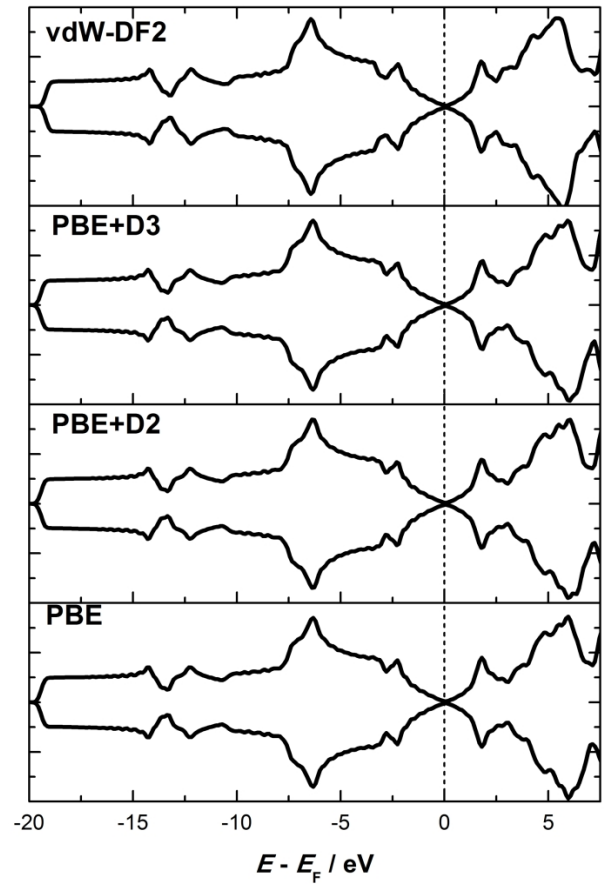

**Figure S2.** DOS curves for the optimized graphene sheets (right). Vertical dashed line indicates Fermi level.

**Table S1.** Preferred adsorption sites for atomic adsorption (t – top site, b – bridge site, h – hollow site, na – no adsorption). If the differences between adsorption sites are under thermal energy at 298 K symbol ● is included.

| H | | | | | | | | | | | | | | | | | He |
|---|---|---|---|---|---|---|---|---|---|---|---|---|---|---|---|---|---|
| t | | | | | element | | | | | | | | | | | | b● |
| t | | | | | PBE | | | | | | | | | | | | h● |
| t | | | | | PBE+D2 | | | | | | | | | | | | h● |
| t | | | | | PBE+D3 | | | | | | | | | | | | h● |
| | | | | | vdW-DF2 | | | | | | | | | | | | |
| Li | Be | | | | | | | | | | | B | C | N | O | F | Ne |
| h | na | | | | | | | | | | | b | b | b | b | t | na |
| h | h | | | | | | | | | | | b | b | b | b | t | t● |
| h | h | | | | | | | | | | | b | b | b | b | t | t● |
| h | b● | | | | | | | | | | | b | b | b | b | t | t● |
| Na | Mg | | | | | | | | | | | Al | Si | P | S | Cl | Ar |
| h | b● | | | | | | | | | | | h | b | h● | b | b● | na |
| h | h● | | | | | | | | | | | h | b | h● | b | t● | h● |
| h | h● | | | | | | | | | | | h | b | h● | b | b● | b● |
| h | b● | | | | | | | | | | | b● | t● | b● | h | b● | h● |
| K | Ca | Sc | Ti | V | Cr | Mn | Fe | Co | Ni | Cu | Zn | Ga | Ge | As | Se | Br | Kr |
| h | h | h | h | h | b● | h | h | h | h | b● | t● | h | b● | b● | h● | t● | b● |
| h | h | h | h | h | b | h● | h | h | h | t● | h● | h | b● | h● | b | b● | h● |
| h | h | h | h | h | b● | h | h | h | h | t● | h● | h | b● | h● | h● | h● | h● |
| h | h | h | h | h | t● | h | h | t | b | h● | h● | b● | t● | b● | t● | b● | h● |
| Rb | Sr | Y | Zr | Nb | Mo | Tc | Ru | Rh | Pd | Ag | Cd | In | Sn | Sb | Te | I | Xe |
| h | h | h | h | h | h● | h | h | h | b | b● | b● | h | t● | b | b● | t● | na |
| h | h | h | h | h | h | h | h | h | b● | h● | h● | b● | b● | h● | b● | t● | h● |
| h | h | h | h | h | b● | h | h | h | b | h● | h● | h● | t● | h● | b● | h● | b● |
| h | h | h | h | h | t● | b● | h | h | b● | h● | t | t● | t● | b● | b● | b● | b● |
| Cs | Ba | La | Hf | Ta | W | Re | Os | Ir | Pt | Au | Hg | Tl | Pb | Bi | Po | At | Rn |
| h | h | h | h | h | h | h | h | b | b | t● | b● | h | t● | h● | h● | h● | na |
| b | b● | h | h | h | h | h | h | b | b | t● | h● | t● | t● | h● | t● | h● | h● |
| h | b● | h | h | h | h | b | h | b | b | t● | h● | h● | t● | b● | t● | h● | b● |
| h | h | h | h | h | b | t | t | b | b | t● | h● | b● | t● | b● | t● | h● | t● |

**Table S2.** Adsorbate-carbon distances (given in Å) for the preferred adsorption sites presented in Table S1.

| H | | | | | | | | | | | | | | | | | He |
|---|---|---|---|---|---|---|---|---|---|---|---|---|---|---|---|---|---|
| 1.126 | | | | | element | | | | | | | | | | | | 3.419 |
| 1.125 | | | | | PBE | | | | | | | | | | | | 3.215 |
| 1.127 | | | | | PBE+D2 | | | | | | | | | | | | 3.286 |
| 1.123 | | | | | PBE+D3 | | | | | | | | | | | | 3.226 |
| | | | | | vdW-DF2 | | | | | | | | | | | | |
| Li | Be | | | | | | | | | | | B | C | N | O | F | Ne |
| 2.230 | na | | | | | | | | | | | 1.842 | 1.527 | 1.462 | 1.459 | 1.561 | na |
| 2.271 | 2.049 | | | | | | | | | | | 1.842 | 1.526 | 1.461 | 1.458 | 1.560 | 2.910 |
| 2.230 | 2.052 | | | | | | | | | | | 1.848 | 1.527 | 1.462 | 1.459 | 1.560 | 2.947 |
| 2.264 | 3.654 | | | | | | | | | | | 1.696 | 1.534 | 1.487 | 1.504 | 1.823 | 2.933 |
| Na | Mg | | | | | | | | | | | Al | Si | P | S | Cl | Ar |
| 2.724 | 3.924 | | | | | | | | | | | 2.567 | 2.189 | 3.770 | 1.918 | 3.206 | na |
| 2.669 | 3.491 | | | | | | | | | | | 2.571 | 2.191 | 3.457 | 1.918 | 3.006 | 3.477 |
| 2.723 | 3.673 | | | | | | | | | | | 2.611 | 2.192 | 3.611 | 1.918 | 3.170 | 3.445 |
| 2.861 | 3.805 | | | | | | | | | | | 2.454 | 2.635 | 3.461 | 3.556 | 3.185 | 3.612 |
| K | Ca | Sc | Ti | V | Cr | Mn | Fe | Co | Ni | Cu | Zn | Ga | Ge | As | Se | Br | Kr |
| 2.960 | 2.708 | 2.439 | 2.331 | 2.333 | 2.374 | 2.516 | 2.109 | 2.101 | 2.114 | 2.191 | 3.855 | 2.630 | 2.430 | 4.079 | 3.699 | 3.115 | 3.888 |
| 2.974 | 2.734 | 2.432 | 2.324 | 2.327 | 2.377 | 2.514 | 2.105 | 2.094 | 2.109 | 2.069 | 3.416 | 2.627 | 2.432 | 3.554 | 2.142 | 3.321 | 3.664 |
| 2.986 | 2.712 | 2.439 | 2.331 | 2.333 | 2.381 | 2.559 | 2.131 | 2.100 | 2.112 | 2.101 | 3.582 | 2.716 | 2.473 | 3.769 | 3.550 | 3.586 | 3.785 |
| 3.099 | 2.882 | 2.567 | 2.435 | 2.456 | 2.517 | 3.823 | 3.498 | 2.214 | 2.014 | 2.483 | 3.783 | 2.688 | 2.696 | 3.577 | 3.282 | 3.378 | 3.774 |
| Rb | Sr | Y | Zr | Nb | Mo | Tc | Ru | Rh | Pd | Ag | Cd | In | Sn | Sb | Te | I | Xe |
| 3.119 | 2.884 | 2.563 | 2.450 | 2.275 | 2.204 | 2.185 | 2.239 | 2.276 | 2.173 | 3.591 | 4.131 | 2.808 | 2.702 | 3.872 | 3.693 | 3.691 | na |
| 3.121 | 2.915 | 2.554 | 2.440 | 2.265 | 2.201 | 2.181 | 2.230 | 2.276 | 2.173 | 3.263 | 3.494 | 2.650 | 2.746 | 3.691 | 3.377 | 3.424 | 3.816 |
| 3.129 | 2.887 | 2.563 | 2.450 | 2.275 | 2.478 | 2.184 | 2.249 | 2.277 | 2.176 | 3.561 | 3.727 | 2.952 | 2.716 | 3.947 | 3.538 | 3.694 | 3.649 |
| 3.276 | 3.128 | 2.776 | 2.554 | 2.447 | 2.796 | 2.973 | 2.365 | 2.200 | 2.313 | 3.430 | 3.644 | 2.882 | 2.941 | 3.711 | 3.684 | 3.660 | 3.624 |
| Cs | Ba | La | Hf | Ta | W | Re | Os | Ir | Pt | Au | Hg | Tl | Pb | Bi | Po | At | Rn |
| 3.259 | 2.964 | 2.671 | 2.401 | 2.348 | 2.220 | 2.208 | 2.233 | 2.072 | 2.097 | 2.451 | 4.095 | 2.939 | 2.881 | 4.032 | 4.443 | 4.172 | na |
| 3.249 | 2.779 | 2.644 | 2.391 | 2.338 | 2.231 | 2.202 | 2.232 | 2.073 | 2.093 | 2.343 | 3.470 | 2.673 | 2.835 | 3.687 | 3.400 | 3.776 | 3.745 |
| 3.275 | 2.978 | 2.676 | 2.402 | 2.348 | 2.238 | 3.374 | 2.258 | 2.077 | 2.098 | 2.489 | 3.583 | 3.016 | 2.901 | 3.822 | 3.547 | 3.900 | 3.693 |
| 3.418 | 3.123 | 2.824 | 2.595 | 2.495 | 2.442 | 3.576 | 3.610 | 2.162 | 2.177 | 3.126 | 3.747 | 3.042 | 2.973 | 3.780 | 3.737 | 4.025 | 3.657 |

**Table S3.** Calculated adsorption energies (in eV) of atoms in the PTE on pristine graphene for the preferred adsorption sites provided in Table S1.

| H | | | | | | | | | | | | | | | | | He |
|---|---|---|---|---|---|---|---|---|---|---|---|---|---|---|---|---|---|
| −0.84 | | | | | element | | | | | | | | | | | | −0.01 |
| −0.91 | | | | | PBE | | | | | | | | | | | | −0.03 |
| −0.88 | | | | | PBE+D2 | | | | | | | | | | | | **−0.04** |
| **−0.91** | | | | | PBE+D3 | | | | | | | | | | | | −0.03 |
| | | | | | vdW-DF2 | | | | | | | | | | | | |
| Li | Be | | | | | | | | | | | B | C | N | O | F | Ne |
| −1.19 | na | | | | | | | | | | | −0.88 | −1.46 | −0.91 | −2.06 | −1.73 | na |
| **−1.56** | −0.10 | | | | | | | | | | | **−1.04** | **−1.59** | **−1.03** | **−2.15** | **−1.83** | −0.04 |
| −1.23 | −0.02 | | | | | | | | | | | −0.96 | −1.52 | −0.98 | −2.12 | −1.79 | −0.03 |
| −1.05 | **−0.16** | | | | | | | | | | | −0.50 | −1.04 | −0.59 | −1.48 | −1.65 | **−0.04** |
| Na | Mg | | | | | | | | | | | Al | Si | P | S | Cl | Ar |
| −0.55 | −0.03 | | | | | | | | | | | −0.88 | −0.44 | −0.00 | −0.66 | −0.79 | na |
| **−0.93** | −0.11 | | | | | | | | | | | **−1.09** | **−0.64** | −0.12 | **−0.83** | −0.92 | −0.08 |
| −0.64 | −0.24 | | | | | | | | | | | −0.99 | −0.55 | −0.14 | −0.80 | −0.93 | −0.09 |
| −0.49 | **−0.16** | | | | | | | | | | | −0.60 | −0.30 | **−0.19** | −0.46 | **−1.04** | **−0.11** |
| K | Ca | Sc | Ti | V | Cr | Mn | Fe | Co | Ni | Cu | Zn | Ga | Ge | As | Se | Br | Kr |
| −0.94 | −0.65 | −1.37 | −1.68 | −1.07 | −0.24 | −0.15 | −0.70 | −1.22 | −1.40 | −0.27 | −0.02 | −0.73 | −0.26 | −0.02 | −0.12 | −0.53 | −0.01 |
| **−1.24** | **−0.91** | **−1.61** | **−1.93** | **−1.32** | **−0.52** | **−0.38** | **−0.98** | **−1.67** | **−1.67** | **−0.55** | −0.17 | **−0.99** | **−0.53** | −0.17 | −0.32 | −0.73 | −0.13 |
| −1.07 | −0.75 | −1.48 | −1.80 | −1.18 | −0.38 | −0.30 | −0.82 | −1.32 | −1.50 | −0.44 | **−0.21** | −0.90 | −0.41 | −0.15 | −0.27 | −0.72 | −0.12 |
| −0.86 | −0.42 | −0.75 | −0.90 | −0.71 | −0.31 | −0.18 | −0.30 | −0.25 | −0.73 | −0.22 | −0.15 | −0.53 | −0.28 | **−0.20** | **−0.37** | **−0.80** | **−0.14** |
| Rb | Sr | Y | Zr | Nb | Mo | Tc | Ru | Rh | Pd | Ag | Cd | In | Sn | Sb | Te | I | Xe |
| −0.95 | −0.60 | −1.46 | −2.02 | −1.36 | −0.15 | −1.14 | −1.60 | −1.76 | −1.11 | −0.03 | −0.02 | −0.57 | −0.15 | −0.01 | −0.06 | −0.35 | nb |
| **−1.32** | **−0.95** | **−1.78** | **−2.36** | **−1.73** | **−0.53** | **−1.53** | **−1.97** | **−2.13** | **−1.47** | **−0.32** | **−0.22** | **−0.95** | **−0.50** | −0.22 | −0.28 | −0.55 | **−0.18** |
| −1.11 | −0.73 | −1.58 | −2.15 | −1.48 | −0.29 | −1.27 | −1.78 | −1.90 | −1.29 | −0.24 | −0.22 | −0.78 | −0.32 | −0.15 | −0.25 | −0.52 | −0.16 |
| −0.88 | −0.49 | −1.01 | −1.14 | −0.94 | −0.27 | −0.29 | −0.78 | −1.00 | −0.82 | −0.20 | −0.16 | −0.53 | −0.28 | **−0.22** | **−0.31** | **−0.59** | −0.16 |
| Cs | Ba | La | Hf | Ta | W | Re | Os | Ir | Pt | Au | Hg | Tl | Pb | Bi | Po | At | Rn |
| −1.11 | −1.05 | −1.93 | −1.49 | −1.57 | −0.52 | −0.05 | −0.89 | −1.33 | −1.46 | −0.13 | −0.02 | −0.56 | −0.14 | −0.01 | −0.05 | −0.26 | na |
| **−2.17** | **−1.95** | **−2.75** | **−2.13** | **−2.16** | **−1.13** | **−0.66** | **−1.49** | **−1.90** | **−2.02** | **−0.69** | **−0.36** | **−0.89** | **−0.53** | **−0.28** | **−0.29** | **−0.50** | **−0.24** |
| −1.28 | −1.08 | −2.07 | −1.62 | −1.72 | −0.74 | −0.33 | −1.13 | −1.54 | −1.76 | −0.36 | −0.25 | −0.82 | −0.34 | −0.16 | −0.25 | −0.45 | −0.17 |
| −1.03 | −0.95 | −1.42 | −0.81 | −0.45 | −0.23 | −0.22 | −0.29 | −0.54 | −1.06 | −0.42 | −0.17 | −0.57 | −0.30 | −0.22 | −0.28 | −0.49 | −0.17 |

**Table S4.** Adsorption energies of selected atoms in the PTE obtained using PAW sets with different electron configurations (t – top site, b – bridge site, h – hollow site)

| element | valence electrons | PBE | | PBE+D2 | | PBE+D3 | | vdW-DF2 | |
|---|---|---|---|---|---|---|---|---|---|
| | | ads. site | $E_{ads}$ / eV | ads. site | $E_{ads}$ / eV | ads. site | $E_{ads}$ / eV | ads. site | $E_{ads}$ / eV |
| Al | 3 | h | −0.91 | h | −1.12 | h | −1.02 | b | −0.61 |
| | 11 | h | −0.88 | h | −1.09 | h | −0.99 | b | −0.60 |
| Si | 4 | b | −0.45 | b | −0.65 | b | −0.57 | t | −0.35 |
| | 12 | b | −0.44 | b | −0.64 | b | −0.55 | t | −0.30 |
| Fe | 8 | h | −0.70 | h | −0.98 | h | −0.82 | t | −0.30 |
| | 14 | h | −0.75 | h | −1.03 | h | −0.87 | t | −0.17 |
| Ni | 10 | h | −1.40 | h | −1.67 | h | −1.50 | b | −0.73 |
| | 18 | h | −1.42 | h | −1.68 | h | −1.51 | b | −0.65 |
| Cu | 11 | b | −0.27 | t | −0.55 | t | −0.44 | t | −0.22 |
| | 17 | b | −0.28 | t | −0.55 | t | −0.44 | t | −0.20 |
| Zn | 12 | t | −0.02 | h | −0.17 | h | −0.21 | h | −0.15 |
| | 20 | t | −0.02 | h | −0.18 | h | −0.21 | h | −0.15 |
| Zr | 4 | h | −2.05 | h | −2.39 | h | −2.18 | h | −1.17 |
| | 12 | h | −2.02 | h | −2.36 | h | −2.15 | h | −1.14 |